\documentclass[amsmath,amssymb,amssymb]{revtex4-1}

\usepackage{graphicx,epsfig,epstopdf}
\usepackage{mathrsfs}
\usepackage{dcolumn}
\usepackage{bm}
\usepackage{natbib}
\begin{document}
\title{Discriminating Cascading Processes in Nonlinear Optics; A QED Analysis Based on Their Molecular and Geometric Origin}
\author{Kochise Bennett$^{a,b}$}
\email{kcbennet@uci.edu}
\author {Vladimir Y. \surname{Chernyak}$^{c,d}$}
\author{Shaul Mukamel$^{a,b}$}
\email{smukamel@uci.edu}
\affiliation{$^a$Chemistry Department, University of California, Irvine, California 92697-2025, USA}
\affiliation{$^b$Department of Physics and Astronomy, University of California, Irvine, California 92697-2025, USA}
\affiliation{$^c$Department of Mathematics, Wayne State University,
656 W. Kirby, Detroit, MI 48202}
\affiliation{$^d$Department of Chemistry, Wayne State University,
5101 Cass Ave, Detroit, MI 48202}
\date{\today}
\begin{abstract}
The nonlinear optical response of a system of molecules often contains contributions whereby the products of lower-order processes in two seperate molecules give signals that appear on top of a genuine direct higher-order process with a single molecule.  These many-body contributions are known as cascading and complicate the interpretation of multidimensional stimulated Raman and other nonlinear signals.  In a quantum electrodynamic (QED) treatment, these cascading processes arise from second-order expansion in the molecular coupling to vacuum modes of the radiation field, i.e., single-photon exchange between molecules, which also gives rise to other collective effects. We predict the relative phase of the direct and cascading nonlinear signals and its dependence on the microsocopic dynamics as well as the sample geometry. This phase may be used to identify experimental conditions for distinguishing the direct and cascading signals by their phase. Higher order cascading processes involving the exchange of several photons between more than two molecules are discussed.  
\end{abstract}
\maketitle

\section{introduction}

Ultrafast nonlinear optical signals in the condensed phase usually use heterodyne detection, whereby the signal interferes with a reference local oscillator beam.  The signal then scales linearly with the number of molecules $N$ and is distinguished by its power dependence in the incoming fileds and its direction (phase matching) \cite{mukbook,boydnonlinear}. Matter information is contained in multipoint correlation functions of the dipole operator.  We will refer to such signals as direct.  While linear spectroscopy reveals the one-photon excitation spectrum of the sample, nonlinear spectroscopies contain multi-photon transitions and can probe dynamics in a controlled way.  For example, in a pump-probe stimulated Raman experiment, a sample is excited by two interactions with an initial light pulse to a vibrational coherence which is then probed by a second pulse following a time delay $T_1$.  This technique probes the vibrational structure of the molecule and is a 3rd-order process in the electric field (and is therefore related to $\chi^{(3)}$, the 3rd-order susceptibility).  An extension of this technique is a 5th-order Raman process in which, following the initial delay $T_1$, a second pair of interactions transfers the sample from one vibrational coherence to another.  The system then evovles freely during a second interpulse delay $T_2$ before being probed.  This more general 5th-order ($\chi^{(5)}$) technique offers the possibility of probing the coupling between participating vibrational modes.
\par
Another type of signal is also generated by a sequential process: some fields interact with one molecule to generate a field which then acts on another molecule, together with the external fields, to finally produce a signal.  Such signals scale as $N^2$ and are known as cascading.  The cascading signals have the same power dependence in the incoming fields and the same phase matching and are thus hard to distinguish from the direct signal \cite{blank1999,golonzka2000,astinov2000diffractive2, astinov2000diffractive2,kubarych2003,li2008two,mehlenbacher2009,bennett2014cascading}.  In neat liquids, cascading dominates the direct signal due to the quadratic \textit{vs.}\ linear scaling in molecular density.  It is then hard to separate the direct signal, which carries a higher level of molecular information.  An important example is the direct 5th-order Raman $\chi^{(5)}$ process which is accompanied by a product of two 3rd-order $\chi^{(3)}$ signals. Separating the two had drawn considerable attention and took several years to recognize \cite{tanimura1993two,tominaga1996fifth,tominaga1996temporally,steffen1997time, steffen1997analysis,TokmakoffFleming,blank1999,golonzka2000, kubarych2003,li2008two}.  This has been the main obstacle for multidimensional Raman spectroscopy (5th-and-higher orders) \cite{mehlenbacher2009}.
\par
A number of ideas have been pursued to separate these undesired cascading signals.  The $N$ \textit{vs.}\ $N^2$ scaling is an obvious way to separate the cascading and direct contributions but in many cases (e.g., 5th-order Raman in neat molecular liquids), it is not possible to vary the molecular density over a sufficiently-large range in order to seperate the $N$- and $N^2$-scaling contributions.  Another idea comes from the fact that, in an infinite homogeneous sample, it follows from Maxwell's equations that the electric field emitted by a polarization is $\frac{\pi}{2}$ phase-shifted relative to the polarization itself.  From a macroscopic perspective, the electric field $\mathbf{E}$ created by a polarization $\bm{\mathcal{P}}$ is given by the Maxwell equation
\begin{align}
\left(\frac{\partial^2}{\partial t^2}-c^2\nabla^2\right)\mathbf{E}(\bm r,t)=-\frac{\partial^2}{\partial t^2}\bm{\mathcal{P}}(\bm r,t)
\end{align} 
which in the frequency domain reads
\begin{align}
\mathbf{E}(\bm k,\omega)=\frac{\omega^2}{\omega^2-k^2c^2+i\eta}\bm{\mathcal{P}}(\bm k,\omega)
\end{align}
where $\eta$ is a positive infinitessimal selecting the appropriate boundary conditions of an outgoing wave.  In the limit of perfect phase matching $\omega^2\to k^2c^2$, we have
\begin{align}
\mathbf{E}(\bm k,\omega)=i\pi\delta(\omega^2-k^2c^2)\omega^2\bm{\mathcal{P}}(\bm k,\omega).
\end{align}
This gives a $\pi/2$ phase shift between the polarization and the resulting field.  We will derive a more general, geometry-dependent result and show under what conditions it reduces to this simple macroscopic relation. The extra emission event (from the source molecule) in a cascading process thus renders the cascading signal out of phase relative to the direct nonlinear signal ($E\propto P-iPP$) \cite{mukbook, li2008two,frostig2015single}.  This property has been parlayed into a number of successfully-implemented techniques to select for the direct 5th-order Raman signal and eliminate the cascading contributions \cite{golonzka2000, blank1999, zhao2011, frostig2015single} (techniques based on polarization-sensitive measurements have also been pursued \cite{gelin2013simple}). 
\par  
In this paper, we extend our earlier treatment of cascading to study this phase shift on a microscopic level \cite{bennett2014cascading}. We show that the $\pi/2$ phase-shift described above occurs in the limit of macroscopic homogeneity only for certain geometries and does not hold for samples consisting of only a few particles. The essence of this geometry-related phase comes from 2D integrations as was described by Feynman for the emitted field by an infinite sheet of dipoles \cite{feynman2013}. After tracing over quantum field degrees of freedom, general expressions for cascading signals are given that are valid in the non-perturbative regime with respect to arbitrary external fields. The result can then be expressed solely as a product of two single-molecule dipole correlation functions and a geometric factor describing the relative positions of the molecules in the sample. This presentation allows the calculation of cascading signals using only the single-molecule calculations familiar from standard treatments of nonlinear spectroscopy and a sum of c-numbers associated with the geometry. This paper extends our previous line of reasoning \cite{bennett2014cascading} which framed cascading signals as a subset of a more general class of vacuum-mediated interactions which includes local-field effects and other higher-order processes.  We then obtained microscopic expressions for the first few orders (in external fields) for these vacuum-mediated effects.  The focus of this paper is the study of the macroscopic continuum limit and a comparison with a microscopic collection of (few) molecules. 
\par

\section{Direct \textit{vs}.\ Cascading Signals}\label{sec:cascading}
In the semiclassical approach to nonlinear optical spectroscopy, a quantum material system interacts with only a few classical modes of the radiation field and the remaining infinite number of modes, which are in their vacuum states, are neglected.  However, taking the vacuum modes into account in a fully quantum electrodynamical (QED) treatement can result in several many-body effects caused by photon exchange, i.e., the emission of a photon by one molecule and its absorption by another.  These effects include the scrambling of the time ordering of incoming short pulses where the free-induction decay produced by one molecule can be long-lived and interacts with another molecule, a local-field $\chi^{(1)}$ effect \cite{mukbook,lozovoy2001cascaded,cundiff2002time,bennett2014cascading}.  Additional local-field effects include corrections to the transmission/reflection of a thin film \cite{benedict1991reflection} and the Rabi oscillations of a quantum dot \cite{paspalakis2006local}.  Nonlinearities are also induced in ensembles of noninteracting harmonic oscillators which are otherwise linear \cite{glenn2015photon}.  Dipole-dipole coupling, responsible for, e.g., Forster resonant energy transfer (FRET) and spontaneous quantum synchronization, is also induced by the exchange of photons \cite{thirunamachandran1980intermolecular, zhu2015synchronization}.  Superradiance, a cooperative spontaneous emission process, is another well-studied effect that finds its origin in the quantum nature of the radiation field \cite{dicke1954coherence}.  Quantum-field effects due to quantum fluctuations of the laser field rather than the unoccupied modes of the electromagnetic field have also been noted in stimulated signals \cite{leon2016genuinely}. This paper analyzes the aforementioned cascading contribution to heterodyne-detected nonlinear spectroscopic signals by treating them as the lowest-order many-body correction due to photon exchange.  
\par
Cascading shares a common origin with other collective effects observed in spectroscopic experiments.  For example, superradiant emission also arises from collective coupling of the constituent molecules of the sample to a photon mode and scales quadratically in the number of molecules \cite{gross1982superradiance}.  However, in superradiance the spontaneously emitted photons (or the associated excited-state population decay) are ordinarily the object of detection while in cascading the exchanged photons are virtual and the relevant experiment is a heterodyne detection with respect to the externally-applied signal (laser) field. The Lamb shift and collective analogues are also due to virtual-photon exchange but these effects are observed as alterations of a material resonance rather than an entirely new signal as in cascading \cite{scully2009collective}.  Forster transfer (discussed more thoroughly in the conclusion) can also be derived perturbatively as a vacuum-mediated interaction.  When the field-matter interaction is treated in the dipole approximation, all of the above depend on the dipole-dipole coupling tensor, giving the derivations a similar flavor \cite{thirunamachandran1980intermolecular, salam}.
\par
We consider a sample made of identical, noninteracting molecules with non-overlapping charge distributions for which the coupling to the radiation field can be treated in the dipole approximation.  The total system dipole operator then takes the form of a sum over molecular dipoles $\hat{\bm{\mathcal{V}}}(\bm{r},t)=\sum_a\hat{\mathbf{V}}(t)\delta(\bm{r}-\bm{r}_a)$.  The material is subjected to a set of classical laser modes $\mathbf{E}_j$ so that the total electric field is
\begin{align}
\hat{\mathbf{E}}(\bm{r},t)\equiv \sum_j \mathbf{E}_j(\bm{r},t)+\hat{\mathbf{E}}_\text{v}(\bm{r},t)
\end{align}
where $\hat{\mathbf{E}}_\text{v}$ is an electric field operator representing the infinitely-many vacuum modes.  
The material system is coupled to this total electric field via the interaction Hamiltonian
\begin{align}\label{eq:Hint}
\hat{H}_{\text{int}}=-\int d\bm{r}\hat{\mathbf{E}}(\bm{r},t)\cdot\hat{\bm{\mathcal{V}}}(\bm{r},t)=-\int d\bm{r}\sum_j\mathbf{E}_j(\bm{r},t)\cdot\hat{\bm{\mathcal{V}}}(\bm{r},t)-\int d\bm{r}\hat{\mathbf{E}}_\text{v}(\bm{r},t)\cdot\hat{\bm{\mathcal{V}}}(\bm{r},t) =\hat{H}_{\text{LM}}+\hat{H}_{\text{vM}}
\end{align}
where the laser modes and the quantum vacuum interact with the material via Hamiltonians $\hat{H}_{\text{LM}}$ and $\hat{H}_{\text{vM}}$ respectively.  
\par
Heterodyne-detected nonlinear spectroscopic signals are given by the rate-of-change of photon number in some signal mode $s$, taken to be in a coherent state.  Commuting the photon number operator with Eq.\ (\ref{eq:Hint}) and integrating over time to get the total photon number change gives the signal
\begin{align}\label{eq:Sdef}
S(\bm k_s,\Lambda)=\Im\bigg[\int d\bm{r}dt\mathbf{E}^*_s(\bm{r},t)\cdot\langle\hat{\bm{\mathcal{V}}}(\bm{r},t)\rangle\bigg] =\Im\left[ \mathbf{E}_s^*(\omega_s)\cdot\bm{\mathcal{P}}(\bm k_s,\omega_s)\right]
\end{align}
where $\mathbf{E}_s$ is the signal field, which we have assumed to have a precisely-defined propagation direction $\hat{\bm k}_s$, and $\Lambda$ stands for the set of parameters defining the classical fields which must be specified to simulate particular signals.  The total polarization is given by the expectation value of the total dipole operator $\langle\hat{\bm{\mathcal{V}}}(\bm{r},t)\rangle=\bm{\mathcal{P}}(\bm{r},t)$, and can be written in terms of time-ordered exponentials
\begin{align}\label{eq:VtimeExp}
\bm{\mathcal{P}}(\bm{r},t)= \text{Tr}\left[\hat{\bm{\mathcal{V}}}(\bm{r},t)\mathcal{T}e^{-i\int_{-\infty}^t d\tau \hat{H}_{-,\text{LM}}(\tau)}e^{-i\int_{-\infty}^t d\tau \hat{H}_{-,\text{vM}}(\tau)}\rho_\text{eq}\right]
\end{align}
where $\rho_\text{eq}$ is the equilibrium field+matter density matrix and ``$-$" subscript stands for the commutator $\hat{H}_-\rho=\hat{H}\rho-\rho\hat{H}$. Expanding Eq.\ (\ref{eq:VtimeExp}) order-by-order in $\hat{H}_\text{LM}$ but to zeroth order in $\hat{H}_\text{vM}$, i.e., neglecting the vacuum modes altogether, yields the standard semiclassical nonlinear optical signals, which we term the direct signal $S_\text{d}$.  All field-matter interactions in the perturbative expansion then occur on a given molecule and the signal is obtained by summing over molecules.  These signals thus scale linearly with the number of molecules in the sample $N$, are proportional to the single-molecule signal suitably averaged, and contain no cooperative many-body contributions.  
\par
The presence of the quantum vacuum modes of the radiation field creates corrections to this picture.  We can systematically generate such corrections by expanding the second exponent in Eq.\ (\ref{eq:VtimeExp}) (the vacuum-mode terms).  Since $\text{Tr}[\hat{a}^{(\dagger)}\vert 0\rangle\langle 0\vert]=0$, the 2nd-order expansion is the lowest non-vanishing correction.  This represents successive emission and re-absorption of a photon by the material.  When the same molecule both emits and absorbs this photon, the result is the lowest-order radiative correction, or one-loop correction, to the energy (Lamb shift with radiative decay \cite{cohen1992atom}).  The terms in which the absorber and emitter are different molecules lead to a transfer of a coherent excitation between the two molecules which is the origin of cascading.  This is depicted diagrammatically in Fig.\ \ref{fig:CascArb}.  
\par
A derivation given in Appendix \ref{app:derivation} gives the total system polarization $\mathcal{P}$ in terms of the individual molecular polarizations $P$ as
\begin{align}\label{eq:Ptotal}
\mathcal{P}^\nu(\bm k_s,\omega)=P^\nu(\bm k_a,\omega)f(\bm k_a-\bm k_s)+\int d\omega_b  \tilde{P}^{\nu\nu_\text{v}}(\bm k_a,\omega_s;-\omega_b)P^{\nu'_\text{v}}(\bm k_b,\omega_b)  G^{\nu_\text{v}\nu'_\text{v}}(\bm k_s-\bm{k}_a,-\bm{k}_b,\omega_b).
\end{align}

where the $\nu$'s denote cartesian coordinates (with summation implicit), $P^{\nu}(\bm k,\omega)$ is the polarization of a single molecule and $ \tilde{P}^{\nu\nu'}(\bm k,\omega;\omega')$ is the polarization of a molecule resulting from a single perturbative interaction with the polarization of another molecule in the sample and arbitrarily-many interactions with the classical fields (defined formally as a dipole correlation function in Eq.\ (\ref{eq:Poldef2})).  In Eq.\ (\ref{eq:Ptotal}), $\bm k_a$ and $\bm k_b$ stand for any linear combination of the set of incoming classical field modes and represent the set of laser modes that interact with molecules $a$ and $b$ respectively in a perturbative expansion. Since we work in the dipole approximation, the $\bm k$-dependence of the polarizations comes only as $\delta$-functions, originating in the spatial phase factor,  that specify the linear combinations. In practice, one must sum over the possible subsets that generate different choices of the $\bm k_a$, $\bm k_b$ (we have omitted integration over $d\bm k_{a}d\bm k_{b}$ for brevity). Finally, $G$ is the photon Green's function defined in $(\bm r,t)$-space in Eq.\ (\ref{eq:Gdef}).
\par
The first term in Eq.\ (\ref{eq:Ptotal}) is the direct nonlinear signal and comes proportional to the sample's form factor 
\begin{align}\label{eq:fdef}
f(\bm{k})\equiv\sum_a e^{i\bm{k}\cdot\bm{r}_a}.
\end{align}
This factor carries information on the position of the molecules in the sample and, in the continuum limit, goes over to the delta function $f(\bm{k})\to(2\pi)^3n\delta(\bm{k})$, where $n=\frac{N}{V}$ is the molecular concentration. This corresponds to momentum conservation and yields the phase matching condition.   The second term in Eq.\ (\ref{eq:Ptotal}) is the cascading signal in which the polarization of molecule $b$ serves as a source, along with the external fields, for the polarization of molecule $a$.  Note that when this is expanded to order $m$ in the classical modes, we will have $\sum_{p+q=m-1}P^{(p)}P^{(q)}$.  The traditional nomenclature is to refer to those terms with $p,q=1$ as local-field corrections and the remainder as cascading (the most familiar being $p=q=3$ but 3,5 cascading has also been discussed \cite{cho2000intrinsic}).  Even though Eq.\ (\ref{eq:Ptotal}) represents a more general class of 2nd-order, vacuum-mediated interactions, we will refer to them collectively as cascading. The key quantity in Eq.\ (\ref{eq:Ptotal}) that connects the polarization emitted by molecule $b$ with the effective field ``felt" by molecule $a$ is the $(\bm k,\omega)$-space photon Green's function
\begin{align}\label{eq:gfunckkw}
G(\bm k,\bm k',\omega)=\sum_{ab}e^{-i(\bm k\cdot\bm r_a+\bm k'\cdot \bm r_b)}G(\bm r_a,\bm r_b,\omega)=\sum_{ab}e^{-i(\bm k\cdot\bm r_a+\bm k'\cdot \bm r_b)} \left(-\nabla^2\delta_{\nu_\text{v}\nu'_\text{v}}+\nabla_{\nu_\text{v}}\nabla_{\nu'_\text{v}}\right)\frac{e^{i\frac{\omega_b}{c}r_{ab}}}{2\pi r_{ab}}
\end{align}
which contains all information about the sample geometry and determines the relative phase of the cascading \textit{vs}.\ direct contributions to the signal.  Moreover, Eq.\ (\ref{eq:Ptotal}) reveals the important point that, in the off-resonant limit where $P(\omega)$ only has the phase imprinted by the external fields, the relative phase of cascading versus direct processes will be controlled entirely by $G(\bm k,\bm{k}',\omega)$. In the limit of a large, homogeneous sample, the spatial dependence of the photon Green's function is reduced to the single variable  $\bm r_a-\bm r_b$, the difference vector, and we have
\begin{align}\label{eq:Glimits}
G^{\nu_\text{v}\nu'_\text{v}}(\bm k_s-\bm{k}_a,-\bm{k}_b,\omega_b)\to f(\bm k_a+\bm k_b-\bm k_s)G^{\nu_\text{v}\nu'_\text{v}}(-\bm{k}_b,\omega_b)\to f(\bm k_a+\bm k_b-\bm k_s)\frac{\omega_b^2}{\omega_b^2- k_b^2c^2+i\eta}\frac{-4\delta_{\nu_\text{v},\nu'_\text{v}}}{3}
\end{align}
where $G(\bm k,\omega)$ directly controls the phase (i.e., if this quantity is real then the cascading process is in phase with the direct process and if it is purely imaginary the two processes are out of phase) and the second relation follows on taking the integral in the continuum limit.  In this regime, the total polarization is thus
\begin{align}\label{eq:Ptotal2}
\mathcal{P}^\nu(\bm k_s,\omega)=P^\nu(\bm k_a,\omega_s)f(\bm k_a-\bm k_s)- i\frac{4\pi nk_bc}{3}\tilde{P}^{\nu\nu_\text{v}}(\bm k_a,\omega_s;\mp k_bc)P^{\nu_\text{v}}(\bm k_b,\pm k_bc)f(\bm k_a+\bm{k}_b-\bm{k}_s).
\end{align}
where we have assumed perfect phase-matching $k_bc=\omega_b$ and both choices of $\pm$ must be summed over.  
\par
Equation (\ref{eq:Ptotal2}) is in the form of the simple macroscopic expression discussed in the introduction that is in common use to understand the cascading contribution to nonlinear signals.  It reveals that the cascading contribution carries an additional factor of molecular density $n$ compared to the direct signal.  Recalling that the form factor also scales linearly with $n$ in the continuum limit, the direct signal is linear while the cascading is quadratic.  The crucial factor of $i$ responsible for the phase shift that is used to filter out cascades is found to originate in the phase-matching condition.  Finally, it is worthwhile to note the linear dependence on $k_bc$ in the cascading contribution, indicating that it will be stronger at higher, such as x-ray, frequencies.  
\par
In making the assumption that $k_bc=\omega_b$, we have neglected the principal value of the denominator in Eq.\ (\ref{eq:Glimits}) which generates a cascading contribution that is \emph{in phase} with the direct signal.  Additionally, the situation can be expected to be different in the case of few-molecule samples or oddly-shaped macroscopic systems.  In the next section, we examine these effects in greater detail.

\begin{figure}
\includegraphics{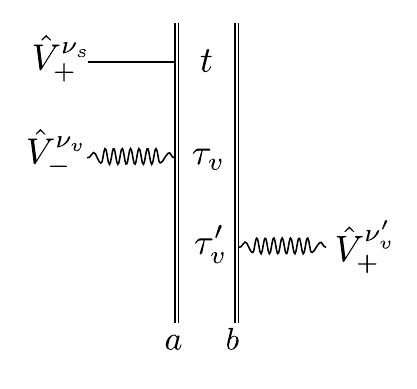}
\caption{This diagram depicts the material quantities relevant for an arbitrary cascading (or, more generally, 2nd-order vacuum-mediated interaction) contribution.  The vertical lines represent the density matrices of molecule $a$ or $b$ and associated propagators.  We use doubled lines to clarify that the propagation is with respect to the full Hamiltonian, including interactions with externally applied fields, and not simply the field-free material propagation.  The straight line intersecting the density matrices of molecule $a$ represents the heterodyne signal field while the wavy lines represent interactions with vacuum modes.  In contrast to the more familiar double-sided Feynman diagrams, we make no distinction between action on the ket or bra.  This is permissible since we work in Liouville space and convenient since it permits us to work in the $+/-$ representation in which the Liouvillian $H_-$ is more compactly written.  This therefore greatly reduces the number of diagrams. Since we work in the $+/-$ representation and without the rotating wave approximation, dressing the interactions with arrows (to indicate positive or negative Fourier components) is unneccesary.  This diagram corresponds to the quantities relevant for equation (\ref{eq:Sexp1}).}
\label{fig:CascArb}
\end{figure}

\section{Sample Geometry and the Phase of the Cascading Signal}\label{sec:geometry}
From the previous section, it is clear that, for the off-resonant response, the $\bm{k}$-space photon Green's function (Eq.\ (\ref{eq:Glimits}) determines the phase of the cascading signal.  In the semiclassical description, the effective electric field emitted by molecule $b$ is given by a phase-shift of the polarization $E\propto iP$.  The extra factor of $i$, which in this semiclassical picture comes from the additional emision event inherent in cascading processes, renders the cascading terms out of phase with the direct nonlinear terms.  This effect has been used to discriminate between the two \cite{golonzka2000, blank1999, zhao2011, frostig2015single}.  In this section, we evaluate the photon Green's function for different geometries and show under what conditions the cascaded signal acquires this well-defined phase shift.
\par
To simplify the analysis, we will separate the summations (integrations) over the positions over the two molecules.  We thus re-write
\begin{align}
&G^{\nu_\text{v}\nu'_\text{v}}(\bm k_s-\bm{k}_a,-\bm{k}_b,\omega_b)=\sum_a e^{i(\bm k_a+\bm k_b-\bm k_s)\cdot\bm r_a} I^{\nu_\text{v}\nu'_\text{v}}(\bm r_a,\bm k_b,\omega_b)\\ 
&I^{\nu_\text{v}\nu'_\text{v}}(\bm r_a,\bm k_b,\omega_b)=\sum_b e^{i\bm k_b\cdot(\bm r_a-\bm r_b)}G^{\nu_\text{v}\nu'_\text{v}}(\bm r_a,\bm r_b,\omega_b),\label{eq:Idef}
\end{align}
where the second summation differs from the discrete Fourier transform of $G$ with respect to $\bm r_b$ only by the spatial phase factor $e^{i\bm k_b\cdot\bm r_a}$.  In the macroscopic homogenous limit, $I$ does not depend on the choice of $\bm r_a$ and we have 
\begin{align}
I^{\nu_\text{v}\nu'_\text{v}}(\bm r_a,\bm k_b,\omega_b)\to I^{\nu_\text{v}\nu'_\text{v}}(\bm k_b,\omega_b)=G^{\nu_\text{v}\nu'_\text{v}}(\bm{k}_b,\omega_b)
\end{align}
in line with Eq.\ (\ref{eq:Glimits}).  Thus, we will primarily be concerned with the evaluation of Eq.\ (\ref{eq:Idef}).

\subsection{A Simplified Treatment for a Scalar Field}\label{sec:response-scalar-contin}
The presence of the dipole coupling tensor $\left(-\nabla^2\delta_{\nu\nu'}+\nabla_{\nu}\nabla_{\nu'}\right)$ in Eq.\ (\ref{eq:gfunckkw}) complicates matters somewhat. For clarity, we first treat the problem by replacing the electromagnetic field, which is a vector gauge field, with a scalar field.  This simplified model still maintains all qualitative features of the original model. The effects will be extended for the original electromagnetic field setting later and we will show that the scalar model describes the relation between the cascaded and direct signals absolutely adequately. 
\par
We assume the following interaction Hamiltonian
\begin{eqnarray}
H_{{\rm int}} = -\int d\bm{r}\varphi(\bm{r})\rho(\bm{r}),
\end{eqnarray}
where $\rho(\bm{r})$ is the scalar polarization, and $\varphi(\bm{r})$ is the scalar field that replaces the full electromagnetic counterpart, so that the Maxwell equations are replaced with the Helmholtz equation ($(\nabla^2+k^2)\phi(\bm r)=0$), whose Green's function (that replaces the Green's function of the Maxwell equations, also known as the Green's function of the electromagnetic field) has the form
\begin{eqnarray}
\label{G-sclalar} G(\bm{r}, \bm{r}'; \omega) = \frac{e^{i\frac{\omega}{c} |\bm{r} - \bm{r}'|}}{|\bm{r} - \bm{r}'|}.
\end{eqnarray}
We will further switch to the uniform continuum limit wherein all properties of the cascaded signal, compared to its direct counterpart are fully contained in the integral

\begin{eqnarray}
\label{define-I}  I(\bm{r}_{a}, \bm k_b,\omega_{b}) = \int_{V} d\bm{r}_b e^{i\bm{k}_{b} \cdot (\bm{r}_b - \bm{r}_{a})}G(\bm{r}_a, \bm{r}_{b}; \omega_{b}),
\end{eqnarray}
where $V$ denotes the integration region, occupied by the sample. We will focus on three special cases. In cases (i) and (ii) there is a poor phase matching, whereas in case (iii) phase matching is good.

Case (i): the region $V$ is a convex $3D$ region with the size large, compared to the wavelength. Introducing an inhomogeneous spherical coordinate system, associated with the convex region $V$, centered at $r_{0}$, with the $z$-axis oriented in the direction of $\bm{k}_{0}$ (see Appendix \ref{sec:measure-sphere}) we recast the integral $I$ in the form
\begin{eqnarray}
\label{I-spherical}  I = \int_{0}^{R}r_b^{2}dr_b \int_{0}^{\pi}\sin\theta d\theta \int_{0}^{2\pi}d\varphi \xi^{3}(\theta, \varphi)(\xi(\theta, \varphi)r_b)^{-1}e^{i\xi(\theta, \varphi)(\frac{\omega_{b}}{c} + k_{b}\cos\theta)r_b},
\end{eqnarray}
where $R$ and $\xi(\theta, \varphi)$, are determined by the shape and size of the region $V$, as well as the position $\bm{r}_{a}$. Assuming a natural condition $k_{b} < \omega_{b}$, integration over $r_a$ can be performed for any values of $\theta$ and $\varphi$, resulting in
\begin{eqnarray}
\label{I-spherical-2}  I = -\int_{0}^{2\pi}d\varphi  \int_{0}^{\pi}\sin\theta d\theta\frac{1}{(\frac{\omega_{b}}{c} + k_{b}\cos\theta)^{2}} + \int_{S^{2}}d\bm{n}\left(\frac{1}{(\frac{\omega_{b}}{c} + \bm{k}_{b} \cdot \bm{n})^{2}}- \frac{iR\xi(\bm{n})}{\frac{\omega_{b}}{c}+ \bm{k}_{b} \cdot \bm{n}} \right)e^{iR\xi(\bm{n})(\frac{\omega_{b}}{c} + \bm{k}_{b} \cdot \bm{n})}.
\end{eqnarray}
The second integral over the unit sphere represents the integral over the boundary of the sample. If the point $\bm{r}_{a}$ is not within the wavelength region from the boundary (which is the typical case), the integrand is a fast oscillating function and the integration can be performed using the saddle point method. In Appendix \ref{sec:saddle-point} we estimate this contribution and demonstrate that the second integral is negligible compared to the first one. Performing the integration in the first term in the r.h.s.\ of Eq.\ (\ref{I-spherical-2}) explicitly, we arrive at
\begin{eqnarray}\label{eq:Gscalar3D}
\label{I-spherical-3}  I = -\frac{4\pi}{\left(\frac{\omega_{b}}{c}\right)^{2} - k_{b}^{2}}.
\end{eqnarray}
This result is natural and just means that, when $\bm{r}_{a}$ is not within a wavelength distance from the boundary, the result is the same as in an infinite medium. Since the contribution of the thin layer in the vicinity of the boundary is minor, the cascaded signal can be computed using the simple expression for $I$, given by Eq.\ (\ref{I-spherical-3}).  Note that we have implicitly assumed poor phase matching, i.e., Eq.\ (\ref{eq:Gscalar3D}) only applies in a principal value sense and neglects the perfect phase matching point $\omega_b=k_bc$.  In the absense of such phase matching, $I$ is clearly real and the cascading signal is in phase with the direct signal.  We will see later that the singular point gives an imaginary, out-of-phase contribution which is dominant under good phase matching.

Case (ii): $V$ is a convex $2D$ region with the size large, compared to the wavelength, such as a molecular monolayer or thin film with thickness smaller than the wavelength.  We can use an inhomogeneous polar coordinate system, resulting in the following expression for the integral $I$
\begin{eqnarray}
\label{I-polar}  I = \int_{0}^{R}r_bdr_b \int_{0}^{2\pi}d\varphi \xi^{2}(\varphi)(\xi(\varphi)r_b)^{-1}e^{i\xi(\varphi)(\frac{\omega_{b}}{c}+ k_{b}\cos\varphi)r_b},
\end{eqnarray}
where in this $2D$ case $\bm{k}_{b}$ naturally denotes the projection of the cascading field wave vector onto to the monolayer/film plane. Performing integration over $r_b$ we obtain
\begin{eqnarray}
\label{I-polar-2}  I =- i \int_{0}^{2\pi}d\varphi \frac{1}{\frac{\omega_{b}}{c}+ k_{b}\cos\varphi} +i \int_{0}^{2\pi}d\varphi \frac{1}{\frac{\omega_{b}}{c}+ k_{b}\cos\varphi}e^{i(\frac{\omega_{b}}{c}+ k_{b}\cos\varphi)a\xi(\varphi)} = I_{0} + I_{1}.
\end{eqnarray}
Neglecting the second contribution coming from a fast-oscillating integral as in the $3D$ case, we obtain
\begin{eqnarray}
\label{I-polar-3}  I = I_{0} =- i \int_{C}\frac{dz}{2\pi iz}\frac{1}{\frac{\omega_{b}}{c} + k_{b}(z + z^{-1})/2} =- i \int_{C}\frac{dz}{2\pi i}\frac{1}{k_{b}z^{2} + \frac{\omega_{b}}{c}z + k_{b}} = \frac{-i}{\sqrt{\left(\frac{\omega_{b}}{c}\right)^{2} - k_{b}^{2}}}.
\end{eqnarray}
This means that, in the off-resonant case where the susceptibilities are real, the direct and cascaded signals are in phase for $3D$, whereas in the $2D$ case there is a $\pi/2$ phase shift between the two signals.

Case (iii): the region $V$ is a cylinder of generically irregular shape located on a reflecting plane. Let $\bm{r} = (x, y, z)$ describing Cartesian coordinates. We assume the region $V$ to be a cylinder of thickness $l$, so that $0 \le z \le l$, where $z = 0$ identifies a reflecting plane. The shape of the cross section of $V$ at $ 0 \le z \le l$ is generally irregular but convex and may depend on $z$. We also assume $\bm{k}_{b} = (0, 0, k_{b})$ to be directed along the $z$-axis, i.e., the cascading wavevector is normal to the plane.  The derivation of this signal follows a similar form to case (ii) but is followed by an integration over $z$.  It is given in detail in Appendix \ref{app:Case3} and results in
\begin{eqnarray}
\label{I-case-iii-phase-matched}  I(\bm{r}_{a},\bm k_b, \omega_{b}) = 2\pi i\frac{c}{\omega_{b}} z_{a} e^{ik_{b}z_{a}}. 
\end{eqnarray}
where we have taken the limit of good phase matching so that  $|\omega_{b} - k_{b}|l \ll 1$ and $z_a$ is not within the wavelength scale from the borders.
\par
To summarize, we note that the $i$ factor in the r.h.s.\ of Eq.\ (\ref{I-case-iii-phase-matched}) appeared due to the $2D$-integration over a cross section (Eq.\ \ref{I-M-of-z})), exactly along the lines of the Feynman's argument \cite{feynman2013}. The third contribution in Eq.\ (\ref{I-case-iii}), which is due to the reflecting surface, vanishes in the good phase matching limit. This tells us that in this limit the effect of the reflecting boundary is negligible, even if it is not perfectly reflecting. Finally, we note that, for a co-linear beam geometry, we have $\bm k_b c=\omega_b$ and this phase-matching condition is automatically satisfied.  In this case, as long as a sufficient number of molecules are involved so as to permit the continuous $2D$-integration over a cross-section of the interaction region, we will obtain a cascading signal exactly out-of-phase with respect to the direct signal.

\subsection{The Full Vector Guage Field}
The previous section demonstrated how to perform the sorts of integrals needed to evaluate $G(\bm r_a, \bm k_b, \omega_b)$ for different geometries utilizing a simplified scalar field.  Restoring the vector nature of the coupling gives for the real space Green's function, 
\begin{align}\label{eq:GvectorReal}
G^{\nu\nu'}(\bm r_a,\bm r_b, \omega_b)=\left(-\nabla^2\delta_{\nu\nu'}+\nabla_{\nu}\nabla_{\nu'}\right)\frac{e^{i\frac{\omega_b}{c}r}}{2\pi r}=\frac{-e^{i\frac{\omega_b}{c}r}}{2\pi r^3}\left[(\delta_{\nu\nu'}-3\hat{r}_{\nu}\hat{r}_{\nu'})(1-i\frac{\omega_b}{c}r)-(\delta_{\nu\nu'}-\hat{r}_{\nu}\hat{r}_{\nu'})\left(\frac{\omega_b}{c}\right)^2r^2\right],
\end{align}
where we have used $r=|\bm r_a-\bm r_b|$ for brevity.  Equation (\ref{eq:GvectorReal}) has terms proportional to $r^{-3}$, $r^{-2}$, and $r^{-1}$.  The last of these terms is purely transverse and, aside from the $\omega_b^2$ factor,  matches the Helmholtz Green's function used as an example in the previous section.  In the limit of an isotropic sample, we have $\hat{r}_{\nu_v}\hat{r}_{\nu'_v}\to\frac{1}{3}\delta_{\nu_v\nu'_v}$ and this $r^{-1}$ term is the only to survive yielding
\begin{align}
G(\bm k_b,\omega_b)=\frac{-4}{3}\frac{\omega_b^2}{\omega_b^2-k_b^2c^2+i\eta}
\end{align}
where we have omitted the cartesian components since the result is proportional to $\delta_{\nu_v\nu'_v}$. This is the equivalent to Eq.\ (\ref{eq:Gscalar3D}) for the vector coupling field and holds under the same geometric assumptions as that equation.  In fact, in the isotropic limit, we may directly obtain the result for the full vector Green's function by simply multiplying the corresponding result obtained above for a scalar field by the factor $\frac{\omega_b^2}{3\pi c^2}$.  More generally, the far-field term is dominant when $\frac{\omega_b r}{c}\gg 1$, corresponding to seperations of greater than $2\pi\lambda$ or roughly 6 times the wavelength of the cascading light.  Thus, for intermolecular seperations much less than this, the static dipole-dipole coupling is dominant while for much larger distances, the long-range $\frac{1}{r}$ term is dominant.  
As demonstrated above, it is this far-field term that leads to a factor of $i$ in the case of a thin sample under good phase matching $\vert k_b-\frac{\omega_b}{c}\vert l\ll 1$.  Note that the $r^{-2}$ term comes with an $i$ factor and hence, when the sample is anisotropic, we may expect that the cascading signal has both in-phase and out-of-phase contributions relative to the direct nonlinear signal.
\par
The limit of an infinite, homogeneous sample can be obtained in a somewhat simpler fashion by integrating over space before handling any vacuum mode summations.  The result is a $\delta$-function selecting the participating vacuum mode and we obtain (see Appendix \ref{app:AltMacro})
\begin{align}\label{eq:Saltfinal}
S_\text{c}(\bm k_s, \Lambda)=-n^2\Im\left[E^{*\nu_s}_s(\omega_s)\int d\omega_bP^{\nu_s\nu_v}(\omega_s;\omega_b)P^{\perp\nu_v}(\bm k_b,\omega_b)\delta(\bm k_a+\bm k_b-\bm k_s)\frac{k_b^2c^2}{\omega_b^2-k_b^2c^2+i\eta}\right],
\end{align}
where we omit factors of $2\pi$ for brevity.  It is thus apparent that the cascading signal consists of a real, principal value part and an imaginary part with a $\delta$-function selecting $\omega_b^2c^2=k_b^2$.  Note that $\bm k_b$ is a linear combination of classical modes interacting with molecule $b$ and $\omega_b$ is the corresponding linear combination of frequencies. Both parameters are therefore externally controlled (up to permutation since which particular classical modes interact with molecule $b$ versus $a$ must clearly be summed over) and we obtain a $\pi/2$ phase shift (factor of $i$) only under perfect phase matching $\omega_b^2=k_b^2c^2$.  As demonstrated in the previous section, this phase-shifted component is dominant for finite samples when the phase-matching is such that $|\omega_b-k_bc|l \ll1$ where $l$ is the optical path length.
\par
Equation (\ref{eq:Saltfinal}) strongly resembles the commonly invoked marcroscopic relation that the cascading field comes as a product of two polarizations $E_\text{c}\propto PP$. Our microscopic derivation reveals the precise sense in which they are related, i.e., through integration over the $(\bm k,\omega)$-space photon Green's function. 
\par
The case of two-molecules is also easily treated and instructive.  Arranging our coordinate system such that the two molecules both lie on the $z$-axis, we obtain the simplified form of the integrated Green's function
\begin{align}
I^{\nu\nu'}(\bm r_a,\bm k_b, \omega_b)=\frac{-\delta_{\nu\nu'}e^{i(\frac{\omega_b}{c}r-\bm k_b\cdot\bm r)}}{2\pi r^3}\left[(1-3\delta_{\nu\nu'z})(1-i\frac{\omega_b}{c}r)-(1-\delta_{\nu\nu'z})\left(\frac{\omega_b}{c}\right)^2r^2\right].
\end{align}
We immediately notice that the Green's function is now a diagonal tensor.  In this two-molecule case, the phase of the cascading signal is sensitively dependent on the distance between molecules and the angle between the cascading beam and the intermolecular axis and comes via the factor $e^{i(\frac{\omega_b}{c}r-\bm k_b\cdot\bm r)}$.  Considering, for example, a co-linear beam geometry perpendicular to the intermolecular axis, we have $\bm k_b\cdot\bm r=0$ and the phase $e^{i\frac{\omega_b}{c}r}$ is a sensitive probe of the intermolecular distance.  Similarly, with a co-linear beam geometry parallel to the intermolecular axis, we obtain a phase of $e^{i(\frac{\omega_b}{c}r-k_b r)}=0$ and the long-range cascading signal is \emph{in phase} with the direct signal.  Finally, we note that the long-range term which usually generates cascading does not contribute to the $G^{zz}$ component.  This means that, if the dipoles are aligned along the axis connecting the two molecules, no long-range cascading takes place.

\section{Conclusions}

Cascading is a vacuum-mediated exchange of coherent polarization between two molecules in a sample.  Being a coherent process, it generates terms with the same phase-matching and scaling with external field amplitudes as direct signals that are ordinarily of more interest since they reveal higher nonlinearities.  In this paper, we have provided a microscopic QED derivation of cascading processes to arbitrary order in the classical modes and connected this to the common macroscopic result obtained via Maxwell's equations.  In particular, we have demonstrated that the $i$ factor used to discriminate cascading from direct signals originates in phase-matching and geometric concerns. In few-molecule samples for example, the cascading signal will generally produce both in-phase and out-of-phase terms of varying dependence on the intermolecular separation vector $\bm r$ and cascading wavevector $\bm k_b$. The uniform integration over the interaction volume cross-section was found to be responsible for this phase-shift, along the same lines as explicated by Feynman for a sheet of dipoles \cite{feynman2013}.  A phase-matching condition $\vert \frac{\omega_b}{c}- k_b\vert l\ll1$, where $l$ is the thickness of the sample, was identified as necessary for the in-phase component of the cascaded signal to vanish. This condition can be achieved in thin films or guaranteed by a co-linear beam geometry. Because they scale quadratically in molecular concentration, the cascading terms often dominate direct nonlinear signals. Additionally, the cascading signal scales linearly with the frequency $\omega_b$ of the emitting polarization.  This implies that cascading processes will become even more dominant at higher frequencies, such as in x-ray Raman experiments, necessitating a thorough understanding of cascading processes for data analysis purposes \cite{mukamel2013multidimensional}.
\par
We are now in a position to appreciate the difference between cascading and fluorescence resonant energy transfer (FRET).  FRET processes are evaluated by taking the square of the Hilbert space amplitude for molecule $b$ to emit a photon then aborbed by molecule $a$ \cite{craig1984molecular,salam2010molecular}.  But this square transition amplitude involves 4 orders in the vacuum mode, and is thus 2 orders higher than cascading.  In fact, in a FRET process the emitting molecule \emph{populates} a photon mode while in cascading the only intermediate photon state is a coherence between the 0- and 1-photon states.  Despite these differences, the derivations in Appendix \ref{app:derivation} have much the same flavor as QED derivations of FRET \cite{craig1984molecular,andrews1989unified,salam2010molecular,scholes2005resonance}, with the principal difference being that cascading depends directly on the dipole coupling tensor rather than its square.  The derivation of Appendix \ref{app:derivation} can thus be extended to account for FRET and other higher-order processes.  At fourth-order, the same as FRET, we also encounter a 3-body cascading contribution that comes as
\begin{align}
\mathcal{P}_{3-\text{body}}^\nu(\bm k_s,\omega_s)&=
\int d\omega_bd\omega_c\tilde{P}^{\nu\nu_\text{v}\mu_\text{v}}(\bm k_a,\omega_s;-\omega_b,\omega_c)P^{\nu'_\text{v}}(\bm k_b,\omega_b)P^{\mu'_\text{v}}(\bm k_c,\omega_c)\\ \notag &\times\sum_{abc} e^{i(\bm k_a+\bm k_b+\bm k_c-\bm k_s)\cdot\bm r_a}I^{\nu_\text{v}\nu'_\text{v}}(\bm r_a,\bm k_b,\omega_b)I^{\mu_\text{v}\mu'_\text{v}}(\bm r_a,\bm k_c,\omega_c)
\end{align}
and will, under phase matching, provide terms that scale cubically in the molecular concentration $n$. Higher-order corrections generate $n$-body cascading terms that follow similarly.  From the perspective of Feynman diagrams, cascading is like a vertex insertion (with four free branches corresponding to the two in-states and two out-states of the participating molecules) while the Lamb shift is the corresponding self-energy insertion.  Both come from vacuum interactions but only the former can be incorporated into a Dyson equation.  The $n$-body cascading terms similarly behave like vertex insertions with $2n$ free branches.
\par
An interesting future extension of this work would be to consider manipulating the cascading signal from a system of molecules embedded in an optical cavity, systems which have drawn recent interest \cite{Hutchison12ac,Shalabney15angew,Simpkins15acsp,kowalewski2016non,kowalewski2016cavity,herrera2016cavity}.  Optical cavities alter the density of electromagnetic field modes from its free-space value, suppressing cascading in all but the cavity mode.  In particular, we note that the field mode participating in a cascading process is determined only for infinite samples (when the mode summation collapses to a $\delta$-functino as in Appendix \ref{app:AltMacro}) while, for few-molecule samples, all vacuum modes participate.  By strongly coupling a few-molecule sample to an optical cavity, we can effectively force the cascading to occur with a particular field mode (the cavity mode). On the other hand, by suppressing the density of field-modes at the cascading wave-vector for a macroscopic system, the cascading signal can be suppressed for arbitrary sample-sizes and beam geometries.  Moreover, molecular coupling to the cavity mode can be tuned by the cavity volume and made much larger than the coupling to the vacuum field.  This would allow for enhancement and control of the cavity-mode cascading.  

\acknowledgements 
The support of the National Science Foundation (grant CHE-1361516) as well as from the Chemical Sciences, Geosciences, and Biosciences division, Office of Basic
Energy Sciences, Office of Science, U.S. Department of Energy through award No. DE-
FG02-04ER15571 is gratefully acknowledged. Support for K.B. was provided by DOE.

\appendix

\section{Microscopic Derivation of Cascading Signals}\label{app:derivation}
In this section, we will derive an expression for the cascading signal from a fully microscopic QED perspective while keeping the perturbative order in the external fields completely arbitrary.  This will allow us to make very general conclusions without considering particulars of the laser fields. 

Expanding equation (\ref{eq:Sdef}) to 2nd order in the vacuum modes of the electric field
\begin{align}
\hat{\mathbf{E}}_\text{v}(\bm{r},t)=i\sum_{\bm{k}_\text{v}\lambda}\sqrt{\frac{2\pi\omega_\text{v}}{\mathcal{V}}}\lbrace \mathbf{\epsilon}^{(\lambda)}(\hat{\bm{k}}_\text{v})e^{i(\bm{k}_\text{v}\cdot\bm{r}-\omega_\text{v}t)}\hat{a}_{\bm{k}_\text{v},\lambda}-\mathbf{\epsilon}^{(\lambda)*}(\hat{\bm{k}}_\text{v})e^{-i(\bm{k}_\text{v}\cdot\bm{r}-\omega_\text{v}t)}\hat{a}^{\dagger}_{\bm{k}_\text{v},\lambda}\rbrace
\end{align}
results in
\begin{align}\label{eq:Sexp1}
&S_{\text{c}}(\Lambda)=\Im\bigg[(i)^2\sum_{ab}\int dt \int_{-\infty}^td\tau_\text{v}\int_{-\infty}^{\tau_\text{v}}d\tau'_\text{v} E^{*\nu_s}_s(\bm{r}_a,t)\langle\hat{V}_+^{\nu_s}(t)\hat{V}_-^{\nu_\text{v}}(\tau_\text{v})\rangle(\bm{r}_a)\langle\hat{V}_+^{\nu'_\text{v}}(\tau'_\text{v})\rangle(\bm{r}_b)\langle\hat{\mathbf{E}}_{\text{v}+}(\bm{r}_a,\tau_\text{v})\hat{\mathbf{E}}_{\text{v}-}(\bm{r}_b,\tau'_\text{v})\rangle_0\bigg]
\end{align}
where the factor of $i^2$ comes from the two orders of expansion in the interaction with the vacuum modes. the $\lbrace\nu\rbrace$ stand for cartesian coordinates coming from the dot products and are implicitly sumed over, and $\langle\dots\rangle_0\equiv \text{Tr}[\dots\vert 0\rangle\langle 0\vert]$ stands for an expectation value taken over the vacuum state. The $-(+)$ subscripts stand for the commutator (anti-commutator) as usual and we have selected the only appropriate contributing terms when performing the initial spatial integrations.  In particular, 3-body terms (in which each of the three dipole operators occur at a different molecule) and that in which the action on molecule $b$ preceedes that on $a$ all vanish since $\text{Tr}[\hat{O}_-\rho]=0$ for all operators $\hat{O}$ and density matrices $\rho$ (this argument is explained in more detail in section II of \cite{bennett2014cascading}). Note however, that 3-body cascading \emph{does} occur when expanding to 4th order in the vacuum modes and cascading 3rd-order processes have been considered in 7th-order nonlinear techniques \cite{cho2000intrinsic}.  Equation (\ref{eq:Sexp1}) is represented diagramatically in Fig.\ \ref{fig:CascArb}. 
\par
It is important to note that the dipole expectation values in Eq.\ (\ref{eq:Sexp1}) includes propagation with respect to the full Hamiltonian (excepting only the vacuum modes).  That is, the externally applied semiclassical modes are still included in the propagators and the expression incorporates all orders of interaction between the matter and laser modes.  Explicitly, we have 
\begin{align}
\langle\hat{V}_+^{\nu_s}(t)\hat{V}_-^{\nu_\text{v}}(\tau_\text{v})\rangle=\text{Tr}\left[\hat{V}_+^{\nu_s}(t)\mathcal{T}\hat{V}_-^{\nu_\text{v}}(\tau_\text{v})e^{-i\int_{-\infty}^td\tau \hat{H}_{-,\text{LM}}(\tau)}\rho(-\infty)\right]
\end{align}
where the time-dependence of the dipole operators is through $\hat{H}_0$, the free material Hamiltonian, and the new interaction Hamiltonian $\hat{H}_\text{LM}$ only includes interactions with the laser fields.  The dependence on molecule position is notated outside the expectation values since all interactions within each expectation value occur on a particular molecule.  Within the dipole approximation, this spatial dependence comes as a simple exponential while the full multi-polar polarization operator has a more general dependence on position.

\par
Equation (\ref{eq:Sexp1}) can be simplified by introducing the polarization of a molecule 
\begin{align}\label{eq:Poldef1}
P^\nu(\bm r,t)=\langle \hat{V}^\nu(t)\rangle(\bm r),
\end{align}
as well as the quantity
\begin{align}\label{eq:Poldef2}
\tilde{P}^{\nu\nu'}(\bm r, t;t')=&i\theta(t-t')\langle\hat{V}_+^{\nu_s}(t)\hat{V}_-^{\nu_\text{v}}(\tau_\text{v})\rangle(\bm r),
\end{align}
which represents the polarization of a molecule in the laser fields due to a perturbative interaction with the polarization of another molecule.
Finally, we identify the photon Green's function 
\begin{align}\label{eq:Gdef}
G^{\nu\nu'}(\bm r_a,\bm r_b, \tau_\text{v}-\tau'_\text{v})=i\theta\left(\tau_\text{v}-\tau_\text{v}'\right)\langle\hat{E}^\nu_{\text{v}+}(\bm{r}_a,\tau_\text{v})\hat{E}^{\nu'}_{\text{v}-}(\bm{r}_b,\tau'_\text{v})\rangle_0,
\end{align}
which comes as a time-ordered, vacuum expectation value of electric field operators.  In terms of these quantities, we may write the cascading signal as
\begin{align}\label{eq:ScascPPGt}
S_\text{c}(\Lambda)=\Im\bigg[\sum_{ab}\int dt d\tau_\text{v}d\tau'_\text{v}E^{*\nu_s}_s(\bm{r}_a,t)\tilde{P}^{\nu_s\nu_\text{v}}(\bm r_a,t;\tau_\text{v})G^{\nu_\text{v}\nu'_\text{v}}(\bm r_a,\bm r_b, \tau_\text{v}-\tau'_\text{v})P^{\nu'_\text{v}}(\bm r_b,\tau'_\text{v})\bigg]
\end{align}
 In appendix \ref{app:field}, we simplify the photon Green's function.  Remaining in the time domain for now, we substitute Eq.\ (\ref{eq:PhotGrt}) to obtain

\begin{align}\label{eq:ScascMicrogen}
S_{\text{c}}(\Lambda)=\Im\bigg[\sum_{ab}\int dt d\tau_v E^{*\nu_s}_s(\bm{r}_a,t)\tilde{P}^{\nu_s\nu_\text{v}}(\bm r_a,t;\tau_\text{v})\left(-\nabla^2\delta_{\nu_\text{v}\nu'_\text{v}}+\nabla_{\nu_\text{v}}\nabla_{\nu'_\text{v}}\right)\frac{1}{2\pi r_{ab}}P^{\nu'_\text{v}}(\bm r_b,\tau_\text{v}-\frac{r_{ab}}{c})\bigg]
\end{align}
where the spatial derivatives (the $\nabla$'s) act on $r_{ab}$, the distance between molecules. This expression is very intuitive; the polarization of molecule $b$, evaluated at the retarded time and adjusted by the action of the dipole coupling tensor and factor of $\frac{1}{r}$, is in place of an external field interaction.  We may make this identification explicit, writing the effective electric field that is felt by molecule $a$ and caused by the polarization emitted from molecule $b$
\begin{align}\label{eq:EcascEff}
E_\text{v}^{\nu_\text{v}}(\bm{r}_a,\tau_v)=\sum_b\left(-\nabla^2\delta_{\nu_\text{v}\nu'_\text{v}}+\nabla_{\nu_v}\nabla_{\nu'_\text{v}}\right)\frac{1}{2\pi r_{ab}}P^{\nu'_\text{v}}(\bm r_b,\tau_\text{v}-\frac{r_{ab}}{c}),
\end{align}
in terms of which the cascading signal is
\begin{align}\label{eq:SmicroRt}
S_{\text{c}}(\Lambda)=\Im\big[\sum_a\int dtd\tau_vE^{*\nu_s}_s(\bm{r}_a,t)E_\text{v}^{\nu_v}(\bm{r}_a,\tau_v)\tilde{P}^{\nu_s\nu_\text{v}}(\bm r_a,t;\tau_\text{v})\big].
\end{align}
Equations (\ref{eq:ScascPPGt})-(\ref{eq:SmicroRt}) have straightforward physical interpretations.  In particular, it is clear that the phase of the cascading signal will depend on the phase of the effective electric field (Eq.\ ($\ref{eq:EcascEff}$)) which in turn depends on the geometry of the sample \textit{via} the summation over positions of molecule $b$.  In physical experiments, the material is subjected to a set of lasers with well-defined propagation vectors $\bm k_j$ and, within the dipole approximation, the real-space dependence of the polarizations will always come as a spatial phase factor $e^{i\sum_j^n\bm k_j\cdot \bm r}$ where $n$ is the order to which the polarization is expanded with respect to the laser-matter interaction.  Thus, the $\bm k$-space polarization will have a delta function setting $\bm k$ to some linear combination of the incoming wavevectors.  In contrast, the spatial structure of the sample can be quite complicated.  We thus transform to $\bm k$-space and, in the interests of brevity, we omit the integrations over $d\bm k_a$ $d\bm k_b$ with the understanding that they will collapse to sums over different choices of $\bm k_a$, $\bm k_b$.  To completely put all geometric dependence on a single term, all else that is required is to change to the frequency domain with respect to the retarded polarization.  We obtain
\begin{align}
S_{\text{c}}(\bm k_s, \Lambda)=\Im\bigg[ E^{*\nu_s}_s(\omega_s)\int  d\omega_b \tilde{P}^{\nu_s\nu_\text{v}}(\bm k_a,\omega_s;-\omega_b)P^{\nu'_\text{v}}(\bm k_b,\omega_b)\sum_{ab}e^{-i(\bm{k}_s-\bm k_a)\cdot\bm{r}_a}e^{i\bm k_b\cdot \bm r_b}G^{\nu_\text{v}\nu'_\text{v}}(\bm r_a,\bm r_b, \omega_b)\bigg]
\end{align}
where we have used the fact that the detected ``signal'' field is  $E^{*\nu_s}_s(\bm{r}_a,t)\to E^{*\nu_s}_s(\omega_s)e^{-i(\bm{k}_s\cdot\bm{r}_a-\omega_st)}$ representing ideal frequency resolution. We explicitly highlight the signal's dependence on this detected mode rather than continuing to include it implicitly in the set of field parameters $\Lambda$. Additionally, we have substituted the spatiotemporal Fourier transforms of Eqs.\ (\ref{eq:Poldef1})-(\ref{eq:Poldef2}).  In terms of the discrete Fourier transform of the photon Green's function
\begin{align}
G^{\nu_\text{v}\nu'_\text{v}}(\bm k_a,\bm k_b, \omega_b)=\sum_{ab}e^{-i\bm{k}_a\cdot\bm{r}_a}e^{-i\bm k_b\cdot \bm r_b}G^{\nu_\text{v}\nu'_\text{v}}(\bm r_a,\bm r_b, \omega_b)
\end{align}
we have 
\begin{align}\label{eq:ScascMicrogen4}
S_{\text{c}}(\bm{k}_s,\Lambda)=&\Im\bigg[E^{*\nu_s}_s(\omega_s)\int d\omega_b  \tilde{P}^{\nu_s\nu_\text{v}}(\bm k_a,\omega_s;-\omega_b)P^{\nu'_\text{v}}(\bm k_b, \omega_b)  G^{\nu_\text{v}\nu'_\text{v}}(\bm{k}_s-\bm k_a,-\bm{k}_b,\omega_b)\bigg]
\end{align}
or
\begin{align}\label{eq:ScascMicrogen4I}
S_{\text{c}}(\bm{k}_s,\Lambda)=&\Im\bigg[E^{*\nu_s}_s(\omega_s)\int d\omega_b  \tilde{P}^{\nu_s\nu_\text{v}}(\bm k_a,\omega_s;-\omega_b)P^{\nu'_\text{v}}(\bm k_b, \omega_b) \sum_a e^{i(\bm{k}_a+\bm{k}_b-\bm{k}_s)\cdot\bm{r}_a} I^{\nu_\text{v}\nu'_\text{v}}(\bm{r}_a,\bm{k}_b,\omega_b)\bigg],
\end{align}
where we have substituted the quantity
\begin{align}\label{eq:IPGF}
I^{\nu_\text{v}\nu'_\text{v}}(\bm r_a,\bm{k}_b,\omega_b)\equiv\sum_{\bm{r}_b} e^{-i\bm{k}_b\cdot\bm{r}_{ab}}G^{\nu_\text{v}\nu'_\text{v}}(\bm r_a,\bm r_b,\omega_b)=\sum_{\bm{r}_b} e^{-i\bm{k}_b\cdot\bm{r}_{ab}}
\left(-\nabla^2\delta_{\nu_\text{v}\nu'_\text{v}}+\nabla_{\nu_\text{v}}\nabla_{\nu'_\text{v}}\right)\frac{e^{i\frac{\omega_b}{c}r_{ab}}}{r_{ab}}
\end{align}
that forms the basis of our discussion of the geometric differences between cascading and direct signals in section \ref{sec:geometry}. Formally, Eq.\ (\ref{eq:ScascMicrogen}) in the time domain and Eqs.\ (\ref{eq:ScascMicrogen4}) or (\ref{eq:ScascMicrogen4I}) in the frequency domain ar our general results.  They give the cascading contribution to a heterodyne-detected signal in terms of the molecular polarization and the photon Green's function, which encodes all geometric information that shapes the cascading signal.  

As an aside, we note that one may wish to remain in the time-domain with respect to molecule $a$, while still handling the geometry more completely than in Eq.\ (\ref{eq:ScascMicrogen}).  This can be accomplished by writing
\begin{align}\label{eq:EcascEffk}
E_\text{v}^{\nu_\text{v}}(\bm r_a,\bm{k}_b,\tau_\text{v})=\int d\omega_b e^{-i\omega_b\tau_\text{v}}\langle\hat{V}_+^{\nu'_\text{v}}(\omega_b)\rangle I^{\nu_\text{v}\nu'_\text{v}}(\bm r_a,\bm{k}_b,\omega_b),
\end{align}
\begin{align}\label{eq:SmicroRw}
S_{\text{c}}(\bm{k}_s,\Lambda)=\Im\big[\sum_a\int dtd\tau_vE^{*\nu_s}_s(\omega_s)e^{i\omega_st}E_\text{v}^{\nu_v}(\bm r_a,\bm{k}_b,\tau_\text{v})\tilde{P}^{\nu_s\nu_\text{v}}(t,\tau_\text{v}) e^{i(\bm{k}_a+\bm{k}_b-\bm{k}_s)\cdot\bm r_a}\big].
\end{align}
where the $\bm r_a$-dependence of $E_\text{v}$ goes away in the homogeneous limit. 

\par
In summary, we first derived an intuitive formula (Eq.\ (\ref{eq:ScascMicrogen})) for the cascading signal based on the polarization of molecule $b$ at the retarded time acting as a source to interact with molecule $a$.  We then obtained a convenient, compact expression (Eq.\ (\ref{eq:ScascMicrogen4I})) for the cascading terms in the heterodyne signal based on the photon Green's function.  It can be readily expanded to any order to obtain cascading corrections to particular nonlinear signals and can also be recast as interaction between an effective electric field and the molecule (Eqs.\ (\ref{eq:EcascEff})-(\ref{eq:SmicroRt}) and (\ref{eq:EcascEffk})-(\ref{eq:SmicroRw})).  The effect of the sample geometry is contained in the exponential phase factor, which approaches the molecule number under good phase matching, and the integral over the photon Green's function $I(\bm r_a,\bm{k}_b,\omega_b)$.  Under off-resonant excitation, the phase of the cascading signal is determined solely by this Green's function.  Finally, we note that everything is currently written in terms of discrete summations over molecular positions but we may take the continuum limit by replacing summation by integration $\sum_{\bm{r}}\to\int d\bm{r}n(\bm{r})$ where $n$ denotes molecular concentration.

\subsection{Alternative Derivation of the Macroscopic Homogeneous Limit}\label{app:AltMacro}
In Appendix \ref{app:derivation}, we performed all possible simplifications, such as vacuum mode summations, bundling all geometric dependence into the photon Green's function and a phase-matching exponential factor.  Such an approach allows full generality in treating different geometries and provides the necessary ingredients for a simulation of cascading processes in few-molecule samples. The resulting expressions can then be applied directly to microscopic geometries or, after converting summations over molecular locations to spatial integrations, various macroscopic geometries and smoothly taken to the infinite limit by integrating over all space.  Some additional insight into this may be obtained by pursuing an alternative derivation in which this spatial integration is performed first, generating $\delta$-functions that determine the participating vacuum mode and collapse the mode sum.  

We begin with 
\begin{align}\label{eq:Sexp2}
S_\text{c}(\Lambda)=&2\Im\bigg[(i)^2\sum_{ab}\sum_{\lbrace\nu\rbrace}\int dt \int_{-\infty}^td\tau_v\int_{-\infty}^{\tau_v}d\tau'_v E^{*\nu_s}_s(\bm{r}_a,t)\langle\hat{V}_+^{\nu_s}(t)\hat{V}_-^{\nu_v}(\tau_v)\rangle(\bm{r}_a)\langle\hat{V}_+^{\nu'_v}(\tau'_v)\rangle(\bm{r}_b)\\ \notag
&\times\sum_{\bm{k}_v\lambda}\frac{2\pi\omega_v}{\mathcal{V}}\lbrace\epsilon_{\nu_v}^{(\lambda)*}(\hat{\bm{k}}_v)\epsilon_{\nu'_v}^{(\lambda)}(\hat{\bm{k}}_v) e^{i(\bm{k}_v\cdot(\bm{r}_a-\bm{r}_b)-\omega_v(\tau_v-\tau'_v)}-\text{c.c}\rbrace\bigg],
\end{align}
obtained from inserting Eq.\ (\ref{eq:Evev2}) into (\ref{eq:Sexp1})
which we then use Eqs.\ (\ref{eq:polsum}) and (\ref{eq:fdef}) to rewrite as
\begin{align}\label{eq:Salt1}
S_\text{c}(\bm k_s, \Lambda)=&2\Im\bigg[(i)^2\sum_{\lbrace\nu\rbrace}\int dt \int_{-\infty}^td\tau_v\int_{-\infty}^{\tau_v}d\tau'_v E^{*\nu_s}_s(\omega_s)e^{i\omega_st}\langle\hat{V}_+^{\nu_s}(t)\hat{V}_-^{\nu_v}(\tau_v)\rangle\langle\hat{V}_+^{\nu'_\text{v}}(\tau'_\text{v})\rangle  \sum_{\bm{k}_v}\frac{2\pi\omega_v}{\mathcal{V}}\left(\delta_{\nu_{v}\nu'_v}-\hat{\bm{k}}_{v}^{\nu_\text{v}}\hat{\bm{k}}_{\text{v}}^{\nu_\text{v}'}\right) \\ \notag
&\times\lbrace f(\bm k_a+\bm k_v-\bm k_s)f(\bm k_b-\bm k_v)e^{-i\omega_\text{v}(\tau_\text{v}-\tau'_\text{v})}-f(\bm k_a-\bm k_\text{v}-\bm k_s)f(\bm k_b+\bm k_\text{v})e^{i\omega_\text{v}(\tau_\text{v}-\tau'_\text{v})}\rbrace\bigg]
\end{align}
where we have substituted the spatial dependence of the dipole expectation values and the signal field $E_s$ as before.  Switching to frequency domain, this can be written as
\begin{align}
S_\text{c}(\bm k_s, \Lambda)=&-2\Im[\sum_{\lbrace \nu\rbrace}\int d\omega_bP^{\nu_s\nu_v}(\omega_s;\omega_b)P^{\nu'_v}(\omega_b)\sum_{\bm k_v}\frac{2\pi\omega_v}{\mathcal{V}} \left(\delta_{\nu_{v}\nu'_v}-\hat{\bm{k}}_{v}^{\nu_v}\hat{\bm{k}}_{v}^{\nu_v'}\right)\\ \notag
&\times\lbrace \frac{f(\bm k_a+\bm k_v-\bm k_s)f(\bm k_b-\bm k_v)}{\omega_b-\omega_v+i\eta}-\frac{f(\bm k_a-\bm k_v-\bm k_s)f(\bm k_b+\bm k_v)}{\omega_b+\omega_v+i\eta}\rbrace
\end{align}
where we have used the definitions of the frequency-domain polarizations of the previous section to simplify the expression. 

In the infinite homogeneous limit, we have $f(\bm{k})\to(2\pi)^3n\delta(\bm{k})$
\begin{align}
S_\text{c}(\bm k_s, \Lambda)=-2(2\pi)^4n^2\Im[E^{*\nu_s}_s(\omega_s)\int d\omega_bP^{\nu_s\nu_v}(\omega_s;\omega_b)P^{\perp\nu_v}(\bm k_b,\omega_b)\delta(\bm k_a+\bm k_b-\bm k_s)\lbrace\frac{k_bc}{\omega_b-k_bc+i\eta}-\frac{k_bc}{\omega_b+k_bc+i\eta}\rbrace]
\end{align}
where we have identified the transverse part of the polarization
\begin{align}
P^{\perp\nu_v}(\bm k_b,\omega_b)=\sum_{\nu'_v}\left(\delta_{\nu_{v}\nu'_v}-\hat{\bm{k}}_{b}^{\nu_v}\hat{\bm{k}}_{b}^{\nu_v'}\right)P^{\nu'_v}(\omega_b).
\end{align}
Finally, we can combine the two terms above to obtain Eq.\ (\ref{eq:Saltfinal})

\section{Field Vacuum Expectation Value}\label{app:field}
In this section, we evaluate the releveant photon Green's function for this problem defined as a time-ordered, vacuum expectation value of electric field operators
\begin{align}\label{eq:Gdef}
G^{\nu\nu'}(\bm r_a,\bm r_b, \tau_\text{v}-\tau'_\text{v})=i\theta\left(\tau_\text{v}-\tau_\text{v}'\right)\langle\hat{E}^\nu_{\text{v}+}(\bm{r}_a,\tau_\text{v})\hat{E}^{\nu'}_{\text{v}-}(\bm{r}_b,\tau'_\text{v})\rangle_0.
\end{align}
Ignoring the prefactors for the moment and expanding first the $\hat{\mathcal{E}}_-$ operator gives
\begin{align}
-i\sum_{\mathbf{k}_v\lambda}\sqrt{\frac{2\pi\omega_\text{v}}{\mathcal{V}}} \text{Tr}[\hat{\mathbf{E}}_{v+}(\mathbf{r}_a,\tau_\text{v})\lbrace \epsilon^{*(\lambda)}(\hat{\mathbf{k}}_\text{v})e^{-i(\mathbf{k}_\text{v}\cdot\mathbf{r}_b-\omega_\text{v}\tau'_\text{v})}\hat{a}^{\dagger}_{\mathbf{k}_\text{v},\lambda}\rho_\text{v}+\epsilon^{(\lambda)}(\hat{\mathbf{k}}_\text{v})e^{i(\mathbf{k}_\text{v}\cdot\mathbf{r}_b-\omega_\text{v}\tau'_\text{v})}\rho_\text{v}\hat{a}_{\mathbf{k}_\text{v},\lambda}]\rbrace
\end{align}
where $\rho_\text{v}\equiv\vert 0\rangle\langle 0\vert$ is the vacuum density matrix and we have kept only non-vanishing terms.  Due to the cyclic invariance of the trace, we may consider only action from the left with respect to the $\hat{\mathbf{E}}_+$ operator that remains.  This results in 
\begin{align}
&2\sum_{\mathbf{k}_\text{v}\lambda}\sum_{\mathbf{k}'_\text{v}\lambda'}\sqrt{\frac{2\pi\omega_\text{v}}{\mathcal{V}}}\sqrt{\frac{2\pi\omega'_\text{v}}{\mathcal{V}}}\bigg[ \epsilon^{*(\lambda)}(\hat{\mathbf{k}}_\text{v})\epsilon^{(\lambda')}(\hat{\mathbf{k}}'_\text{v})e^{i(\mathbf{k}'_\text{v}\cdot\mathbf{r}_a-\omega'_\text{v}\tau_\text{v})}e^{-i(\mathbf{k}_\text{v}\cdot\mathbf{r}_b-\omega_\text{v}\tau'_\text{v})}\text{Tr}[ \hat{a}_{\mathbf{k}'_\text{v},\lambda'}\hat{a}^{\dagger}_{\mathbf{k}_v,\lambda}\rho_\text{v}]\\ \notag & -\epsilon^{(\lambda)}(\hat{\mathbf{k}}_v)\epsilon^{*(\lambda')}(\hat{\mathbf{k}}'_\text{v})e^{i(\mathbf{k}_\text{v}\cdot\mathbf{r}_b-\omega_\text{v}\tau'_\text{v})}e^{i(\mathbf{k}'_\text{v}\cdot\mathbf{r}_a-\omega'_v\tau_v)}\text{Tr}[\hat{a}^\dagger_{\mathbf{k}'_\text{v},\lambda'}\rho_\text{v}\hat{a}_{\mathbf{k}_\text{v},\lambda}]\bigg].
\end{align}
The basic commutation relation then gives $\text{Tr}[ \hat{a}_{\mathbf{k}'_v,\lambda'}\hat{a}^{\dagger}_{\mathbf{k}_v,\lambda}\rho_v]=\text{Tr}[\hat{a}^\dagger_{\mathbf{k}'_v,\lambda'}\rho_v\hat{a}_{\mathbf{k}_v,\lambda}]=\delta_{\mathbf{k}'_v\mathbf{k}_v}\delta_{\lambda'\lambda}$ and we obtain the expression
\begin{align}\label{eq:Evev2}
\langle\hat{\mathbf{E}}_{v+}(\mathbf{r}_a,\tau_v)\hat{\mathbf{E}}_{v-}(\mathbf{r}_b,\tau'_v)\rangle_0=2\sum_{\mathbf{k}_v\lambda}\frac{2\pi\omega_v}{\mathcal{V}}\epsilon^{*(\lambda)}(\hat{\mathbf{k}}_v)\epsilon^{(\lambda)}(\hat{\mathbf{k}}_v)\lbrace e^{i(\mathbf{k}_v\cdot(\mathbf{r}_a-\mathbf{r}_b)-\omega_v(\tau_v-\tau'_v))}-e^{-i(\mathbf{k}_v\cdot(\mathbf{r}_a-\mathbf{r}_b)-\omega_v(\tau_v-\tau'_v))}\rbrace
\end{align} 
To simplify, we perform the polarization sum and change the vacuum mode summation to an integration (see \cite{craig1984molecular,salam2010molecular})
\begin{align}\label{eq:polsum}
&\sum_\lambda \mathbf{\epsilon}_{\nu_v}^{(\lambda)*}(\hat{\bm{k}}_v)\mathbf{\epsilon}_{\nu'_v}^{(\lambda)}(\hat{\bm{k}}_v)=\delta_{\nu_{v}\nu'_v}-\hat{\bm{k}}_{v}^{\nu_v}\hat{\bm{k}}_{v}^{\nu_v'}\\ 
&\frac{1}{\mathcal{V}}\sum_{\bm{k}_v}\to\int\frac{d\omega_vd\Omega_v\omega_v^2}{(2\pi c)^3}
\end{align}
after which the integration over solid angle can be carried out
\begin{align}
\int d\Omega_v\left(\delta_{\nu_v\nu'_v}-\hat{\bm{k}}_{v}^{\nu_v}\hat{\bm{k}}_{v}^{\nu'_v}\right)e^{\pm i\bm{k}_v\cdot\bm{r}}= \left(-\nabla^2\delta_{\nu_v\nu'_v}+\nabla_{\nu_v}\nabla_{\nu'_v}\right)\frac{\sin{k_vr}}{k_v^3r} \label{eq:dipdip}
\end{align}
which then allows us to easily perform the $\omega_v$ integration
\begin{align}
\int d\omega_v\sin{\left(\omega_v\frac{r_{ab}}{c}\right)}\big[e^{-i\omega_v(\tau_v-\tau'_v)}-e^{i\omega_v(\tau_v-\tau'_v)}\big]=i\pi\big[\delta(\tau'_v-\tau_v-\frac{r_{ab}}{c})-\delta(\tau'_v-\tau_v+\frac{r_{ab}}{c})\big].
\end{align}
Combining this with Eq.\ (\ref{eq:Gdef}) yields
\begin{align}\label{eq:PhotGrt}
G^{\nu\nu'}(\bm r_a,\bm r_b, \tau_v-\tau'_v)=\left(-\nabla^2\delta_{\nu\nu'}+\nabla_{\nu}\nabla_{\nu'}\right) \frac{\delta(\tau'_v-\tau_v+\frac{r_{ab}}{c})}{2\pi r_{ab}}.
\end{align}
with temporal Fourier transform
\begin{align}
G^{\nu\nu'}(\bm r_a,\bm r_b,\omega)=\int d(\tau_v-\tau'_v )e^{i\omega(\tau_v-\tau'_v)}G^{\nu\nu'}(\bm r_a,\bm r_b, \tau_v-\tau'_v)=\left(-\nabla^2\delta_{\nu\nu'}+\nabla_{\nu}\nabla_{\nu'}\right) \frac{e^{i\omega\frac{r_{ab}}{c}}}{2\pi r_{ab}}
\end{align}
where acting the differential operators results in Eq.\ (\ref{eq:GvectorReal}).

\section{Cylindrical Geometry}\label{app:Case3}
In this section, we describe the derivation of $I$ for the case of a cylindrical geometry as discussed in Section \ref{sec:geometry}, case (iii). The field Green function in this case has a form
\begin{eqnarray}
\label{G-reflecting}  G(\bm{r}_{a}, \bm{r}_b; \omega_{b}) &=& \frac{1}{\sqrt{(x_{a} - x_b)^{2} + (y_{a} - y_b)^{2} + (z_{a} - z_b)^{2}}}e^{i\frac{\omega_{b}}{c}\sqrt{(x_{a} - x_b)^{2} + (y_{a} - y_b)^{2} + (z_{a} - z_b)^{2}}} \nonumber \\ &-& \frac{1}{\sqrt{(x_{a} - x_b)^{2} + (y_{a} - y_b)^{2} + (z_{a} + z_b)^{2}}}e^{i\frac{\omega_{b}}{c}\sqrt{(x_{a} - x_b)^{2} + (y_{a} - y_b)^{2} + (z_{a} + z_b)^{2}}} \nonumber \\ &=& G_{M}(\bm{r}_{a}, \bm{r}_b; \omega_{b}) + G_{S}(\bm{r}_{a}, \bm{r}_b; \omega_{b}),
\end{eqnarray}
where $G_{M}$ is the field Green function in the complete space $\mathbb{R}^{3}$, and $G_{S}$ is the correction associated with the surface $S$, determined by the condition $z = 0$, so that $G$ is the complete Green function in the half-space $z \ge 0$, with the reflecting boundary conditions. The representation of Eq.~(\ref{G-reflecting}) represents the integral $I$ of Eq.~(\ref{define-I})
\begin{eqnarray}
\label{define-I-split}  I(\bm{r}_{a},\bm k_b, \omega_{b}) = I_{M}(\bm{r}_{a},\bm k_b, \omega_{b}) + I_{S}(\bm{r}_{a},\bm k_b, \omega_{b})
\end{eqnarray}

We start with computing $I_{M}$; the second integral $I_{S}$ is computed in a similar fashion. Fixing $z$ we compute the integral over the corresponding cross section
\begin{eqnarray}
\label{I-M-of-z}  I(z_b; \bm{r}_{a},\bm k_b, \omega_{b}) = e^{ik_{b}(z_b-z_a)} \int_{S_{z}} dxdy G_{M}(\bm{r}_{a}, x_b, y_b, z_b; \omega_{b}), \;\;\; S_{z} = \{(x, y)\; | \; (x, y, z) \in V\}
\end{eqnarray}
by using inhomogeneous polar coordinates with the center in $(x_{a}, y_{a})$ followed by integration over the inhomogeneous radius, resulting in
\begin{eqnarray}
\label{I-M-of-z-2}  I_{M}(z_b; \bm{r}_{a},\bm k_b, \omega_{b}) = 2\pi i\frac{c}{\omega_{b}}e^{ik_{b}(z_b-z_a) + i\frac{\omega_{b}}{c}|z_b- z_{a}|} - i\frac{c}{\omega_{b}}e^{ik_{b}(z_b-z_a)}\int_{0}^{2\pi}d\varphi e^{i\frac{\omega_{b}}{c}\sqrt{(R\xi(\varphi))^{2} + (z_{0} - z)^{2}}}.
\end{eqnarray}
The integration is performed in a similar way to how it has been carried out for the $2D$ case, using the identity
\begin{eqnarray}
\label{I-M-of-z-2-a}  \frac{\xi^{2}r_bdr_b}{\sqrt{(\xi r_b)^{2} + (z_{a} - z_b)^{2}}} = d\sqrt{(\xi r_b)^{2} + (z_{a} - z_b)^{2}},
\end{eqnarray}
and the fact that the projection of $\bm{k}_{b}$ onto the plane is $0$. For the surface contribution we obtain in a similar way
\begin{eqnarray}
\label{I-S-of-z-2}  I_{S}(z_b; \bm{r}_{a},\bm k_b, \omega_{b}) = 2\pi i\frac{c}{\omega_{b}}e^{ik_{b}(z_b+z_a) + i\frac{\omega_{b}}{c}|z_{a} + z_b|} - i\frac{c}{\omega_{b}}e^{ik_{b}(z_b+z_a)}\int_{0}^{2\pi}d\varphi e^{i\frac{\omega_{b}}{c}\sqrt{(R\xi(\varphi))^{2} + (z_a + z_b)^{2}}}.
\end{eqnarray}
By a similar to case (ii) argument, which involves saddle-point approximation for computing a fast oscillating integral, the second contribution in the r.h.s. of Eq.~(\ref{I-M-of-z-2}) and (\ref{I-S-of-z-2}) can be neglected resulting in
\begin{eqnarray}
\label{I-of-z-2-approx}  I_{M}(z_b; \bm{r}_{a},\bm k_b, \omega_{b}) &=&2\pi i\frac{c}{\omega_{b}}e^{ik_{b}(z_b-z_a) + i\frac{\omega_{b}}{c}|z_b-z_{a}|} \nonumber \\ I_{S}(z_b; \bm{r}_{a}; \omega_{b}, V) &=&2\pi i\frac{c}{\omega_{b}}e^{ik_{b}(z_b+z_a) + i\frac{\omega_{b}}{c}|z_b+z_{a}|}.
\end{eqnarray}

Finally performing integration over $z$ we obtain
\begin{eqnarray}
\label{I-case-iii} I(\bm{r}_{a},\bm k_b, \omega_{b}) &=& 2\pi i\frac{c}{\omega_{b}}\bigg[e^{i\frac{\omega_{b}}{c}z_{a}} e^{\frac{i(k_{b} - \frac{\omega_{b}}{c})z_{a}}{2}}z_{a}{\rm sinc}\left(\frac{(\frac{\omega_{b}}{c} - k_{b})z_{a}}{2}\right) \nonumber \\ &+&e^{-i\frac{\omega_{b}}{c}z_{a}}  e^{\frac{i(k_{b} +\frac{\omega_{b}}{c})(l + z_{a})}{2}}(l - z_{a}){\rm sinc}\left(\frac{(\frac{\omega_{b}}{c}+ k_{0})(l - z_{0})}{2}\right) \nonumber \\ &+& e^{i\frac{\omega_{b}}{c}z_{a}} e^{\frac{i(k_{b} - \frac{\omega_{b}}{c})z_{a}}{2}}z_{a}{\rm sinc}\left(\frac{(\frac{\omega_{b}}{c} - k_{b})z_{a}}{2}\right)\bigg],
\end{eqnarray}
where we have expressed the answer in terms of the sinc function
\begin{eqnarray}
\label{sinc-function} {\rm sinc}(x) = \frac{\sin x}{x},
\end{eqnarray}
with the properties ${\rm sinc}(x) \approx 1$, for $|x| \ll 1$, and $|{\rm sinc}(x)| \le |x|^{-1}$.
In the case $|\omega_{b} - k_{b}|l \ll 1$ of a good phase matching and if $z_a$ is not within the wavelength scale from the borders, the first contribution is dominating, so that we reproduce a well-known result
\begin{eqnarray}
\label{I-case-iii-phase-matched}  I(\bm{r}_{a},\bm k_b, \omega_{b}) = 2\pi i\frac{c}{\omega_{b}} z_{a} e^{ik_{b}z_{a}}. 
\end{eqnarray}

\section{Integration Measure for Inhomogeneous Polar Coordinates}
\label{sec:measure-polar}

In this section we compute the integration measure for inhomogeneous polar coordinates. Inhomogeneous in this context mans that the radius depends on the angle. Formally, let $\xi(\varphi)$ be a positively-defined function of angle that has a property that the region $\{\bm{r} \in \mathbb{R}^{2}\; | \; r(\bm{r}) \le \xi(\varphi\bm{r}) \}$ is convex, where $r(\bm{r})$ and $\varphi(\bm{r})$ are the standard polar coordinates of $\bm{r}$. The inhomogeneous polar coordinates are defined by
\begin{eqnarray}
\label{polar-inhom} r_{x} = r \xi({\varphi})\cos\varphi, \;\;\; r_{y} = r \xi({\varphi})\sin\varphi.
\end{eqnarray}
It is easy to demonstrate that the inhomogeneous polar coordinates is a well-defined coordinate system for the region defined by the condition $r < r_{0}$ for any $r_{0}$, provided the convexity condition is satisfied. A reverse statement is also true, i.e., for any connected convex region and any reference point in it one can (and very easily) identify $r_{0}$ and $\xi(\varphi)$ so that the region is defined by the condition $r < r_{0}$ in the inhomogeneous polar coordinate system, centered in the reference point.

To compute the integration measure we first differentiate
\begin{eqnarray}
\label{polar-inhom} dr_{x} &=& \xi\cos\varphi dr + r\left(\frac{\partial \xi}{\partial \varphi}\cos\varphi - \xi\sin\varphi\right)d\varphi \nonumber \\ dr_{y} &=& \xi\sin\varphi dr + r\left(\frac{\partial \xi}{\partial \varphi}\sin\varphi + \xi\cos\varphi\right)d\varphi
\end{eqnarray}
and then apply the concept of the wedge (i.e., antisymmetric) product to obtain
\begin{eqnarray}
\label{polar-inhom-2} dr_{x} \wedge dr_{y} &=& r\xi\cos\varphi dr \wedge r\left(\frac{\partial \xi}{\partial \varphi}\sin\varphi + \xi\cos\varphi\right)d\varphi \nonumber \\ &+& r\left(\frac{\partial \xi}{\partial \varphi}\cos\varphi - \xi\sin\varphi\right)d\varphi \wedge \xi\sin\varphi dr = r\xi^{2}(\varphi) dr \wedge d\varphi
\end{eqnarray}
where in the derivation we have used the antisymmetric properties of the wedge product, namely $dr \wedge dr = d\varphi \wedge d\varphi = 0$ and $d\varphi \wedge dr = -dr \wedge d\varphi$. This provides a very simple result for the integration measure
\begin{eqnarray}
\label{polar-inhom-3} dr_{x}dr_{y}  = \xi^{2}(\varphi) r dr d\varphi,
\end{eqnarray}
in particular the integration measure does not contain derivative of the function $\xi(\varphi)$ that describes the inhomogeneity of the coordinate system. We will refer to the above property as the homogeneous property of the inhomogeneous coordinate system.

\section{Integration Measure for Inhomogeneous Spherical Coordinates}
\label{sec:measure-sphere}

An inhomogeneous spherical coordinate system is introduced in a similar way:
\begin{eqnarray}
\label{shere-inhom} r_{x} = r \xi({\theta, \varphi}) \sin\theta \cos\varphi, \;\;\; r_{y} = r \xi(\theta, {\varphi})\sin\theta sin\varphi, \;\;\; r_{z} = r \xi({\theta, \varphi}) \cos\theta.
\end{eqnarray}
Obviously, the inhomogeneous spherical system satisfies the same convexity properties as the inhomogeneous polar system. A much more tedious computation, similar to the one for the polar system, presented in Eq.~(\ref{polar-inhom-2}), provides a very simple result
\begin{eqnarray}
\label{sphere-inhom-3} dr_{x}dr_{y}r_{z}  = \xi^{3}(\theta, \varphi) r^{2}\sin\theta dr d\theta d\varphi = \xi^{3}(\bm{n})r^{2}drd\bm{n},
\end{eqnarray}
which shows that the measure is also homogeneous. This means that the homogeneity of an inhomogeneous measure takes place in all dimensions, and there is a general geometrical argument that derives Eq.~(\ref{sphere-inhom-3}) for all dimensions that bypasses tedious computations.

\section{Estimates for the Boundary Contributions Using the Saddle-Point Approximation}
\label{sec:saddle-point}

We start with the $2D$ case Denoting in Eq.~(\ref{I-polar-2})
\begin{eqnarray}
\label{define-S} S(\varphi) = (\omega_{0} + k_{0}\cos\varphi)a\xi(\varphi)
\end{eqnarray}
we obtain in the saddle-point approximation
\begin{eqnarray}
\label{I-polar-saddle}  I_{1} = \frac{1}{\sqrt{2\pi \lambda_{0}}}  \frac{1}{\omega_{0} + k_{0}\cos\varphi_{0}}e^{i(\omega_{0} + k_{0}\cos\varphi_{0})a\xi(\varphi_{0})}, \;\;\; \left(\frac{dS(\varphi)}{d\varphi}\right)_{\varphi = \varphi_{0}} = 0, \;\;\; \lambda_{0} = \left(\frac{d^{2}S(\varphi)}{d\varphi^{2}}\right)_{\varphi = \varphi_{0}}.
\end{eqnarray}
Estimating $\lambda_{0} \sim a\omega_{0} \gg 1$ we see that $I_{1}$ is small compared to $I_{0}$ by a factor $\sim (\sqrt{a\omega_{0}})^{-1}$.


%

\end{document}